\documentstyle[12pt]{article}
\def\DB{\bar D}
\def\XB{\bar X}
\def\YB{\bar Y}
\def\ZB{\bar Z}
\def\DPB{{\bar D}_{+}}
\def\DMB{{\bar D}_{-}}

\def\DM{D_{-}}
\def\DB{{\bar D}}
\def\ib{{\bar i}}
\def\jb{{\bar j}}

\def\pa{\partial}
\def\th{\theta}
\def\thb{\bar \theta}
\def\EE{\begin{equation}}
\def\EN{\end{equation}}
\def\EEN{\begin{eqnarray}}
\def\ENN{\end{eqnarray}}
\def\SATA{\frac{\th_{12}\thb_{12}}{Z_{12}}}
\def\SATB{\frac{\th_{12}\thb_{12}}{(Z_{12})^{2}}}
\def\SATC{\frac{\sqrt{-1}}{4}\frac{\th_{12}}{Z_{12}}}
\def\SATD{\frac{\sqrt{-1}}{4}\frac{\thb_{12}}{Z_{12}}}
\begin{document}
\begin{titlepage}
%%%%%%%%%%% this is the preprint number %%%%%%%%%%%%%%
%\begin{flushright}
%	UTHEP-260	\\
%	July, 1993	\\
%
%\end{flushright}
%%%%%%%%%%%%%%%%%%%%%%%%%%%%%%%%%%%%%%%%%%%%%%%%%%%%%%%%
\vspace{5mm}

\begin{center}
{\LARGE
$N=2$ super $W$ algebra     \\
in half-twisted Landau-Ginzburg model}  \\
\vspace{3cm}
{\large Kenji Mohri} \\
%\vspace{2cm}
%\large
%{\name 	Kenji Mohri}

\vspace{1cm}

{\it
Institute of Physics, University of Tsukuba ,Ibaraki  305, Japan}
\end{center}
\begin{abstract}
We investigate  $N=2$ extended superconformal symmetry,
 using the half-twisted Landau-Ginzburg models. The first example is the
$D_{2n+2}$ -type minimal model. It has been conjectured that this model has a
spin $n$ super $W$ current. We checked this  by the direct computations of
 the BRS cohomology class up to $n=4$.
We observe for $n\le 3$ the super W currents  generate the ring  isomorphic to
 the chiral ring of the model with respect to the classical product.
We thus  conjecture that this isomorphism holds for any $n$.
The next example is $ CP_{n}$ coset model.
In this case we find a sort of  Miura transformation which gives the simple
formula for the super W currents of spin \{1,2,...,n\} in terms of
the chiral superfields. Explicit form of the super W currents and
 their Poisson brackets are obtained for $CP_{2},CP_{3}$ case.
We  also conjecture  that as long as  the classical product
is concerned,
these super W  currents  generate the ring  isomorphic to the chiral ring
of the model and this is checked for $CP_2$ model.
\end{abstract}
\end{titlepage}

%\vspace{1.5cm}
%\large

	%ABSTRACT

%\normalsize
\baselineskip=0.8cm

\baselineskip=0.8cm
\newpage
\section{Introduction}
$N=2$ superconformal field theories have been intensively studied in
recent years \cite{BoKeFr} \cite{VaWa},\cite{LeVaWa}, \cite{KaSu}
,\cite{EgKaMiYa} because of its direct relevance to
the superstring compactification \cite{Ge1} \cite{GrVaWa} \cite{Va2},
topological field theory \cite{EgYa}
and supersymmetric integrable systems \cite{FeMaVaWa}, \cite{FeLeMaWa}
\cite{NoMo}, \cite{EHo} \cite{KoMoNo} \cite{It1}.
The large class of $N=2$ superconformal theories can be realized
as the  infra-red fixed point of
Landau-Ginzburg models \cite{VaWa} \cite{Mar}.
Thus it is important to investigate  the structure of $N=2$
superconformal algebras in the framework of the Landau-Ginzburg description.
To do so the apparent difficulty  is that the action of the Landau-Ginzburg
model does not have conformal invariance.
However Witten has  shown quite recently that under
the operation of half-twist \cite{W3}
the Landau-Ginzburg model turns out to be left-moving conformally invariant
\cite{W2}.
This enable us to extract the essential information about the
$N=2$ superconformal algebra realized  at the infra-red fixed point of
  the Landau-Ginzburg model. \\
In this paper our purpose is to  extend the analysis of Witten
to the multi-variable case and to analyse the classical aspect of
$N=2$ super $W$ symmetry.
In section 2 we clarify the twist operation in the Landau-Ginzburg model.
In particular it is argued that  the Landau-Ginzburg model with the
quasi-homogeneous superpotential $W$ admits the half-twist operation,
so that it gets the conformal symmetry.
In section 3 we treat the $D_{2n+2}$-type minimal model which consists of two
Landau-Ginzburg fields for lower $n$s.
It is shown that the $N=2$ algebra of this model is extended
by an operator of spin $n$.
Moreover we find that super W currents generate a ring isomorphic
to the chiral ring of the model under the classical product.
  In section 3 the $CP_{n}$ coset model is analysed.
We point out that there exists a simple transformation law
which yields the super $W$ currents directly from the chiral superfields.
%Here we find the simple transformation law from chiral super fields
%to super W currents.
The explicit form of the super W currents and their Poisson brackets
are obtained  for $CP_{2}$ and $CP_{3}$ cases.
It is conjectured that in these examples also the super W currents
genetate a ring isomorphic to the chiral ring under the classical product,
and this is proved for $CP_2$ model.
In appendix we collect some useful classical and quantum formulas
for the Landau-Ginzburg model.

\newpage

\section{Twists of Landau-Ginzburg models}
\setcounter{equation}{0}
\subsection{Preliminaries}
The Landau-Ginzburg model is an $N=2$ super field theory described by
{\it n} chiral superfields. The  Lagrange density is
\EEN
L& =& 4\Phi^{i}{\bar \Phi}^{\bar i}\mid_{\th^{4}}-W(\Phi^{i})\mid_{\th^{2}}
-{\bar W}({\bar \Phi}^{\bar i})\mid_{{\bar \th}^{2}} \nonumber \\
&=&\int d\th^{-}d\th^{+}d{\bar \th}^{+}d\thb^{-}\Phi^{i}{\bar \Phi}^{\bar i}
-\frac{1}{2}\int d\th^{-}d\th^{+}W(\Phi^{i})
-\frac{1}{2}\int d\th^{+}d\th^{-}{\bar W}({\bar \Phi}^{\bar i})\nonumber \\
&=& 4(2\sqrt{-1}{\bar \psi}_{-}^{\bar i}\pa_{+}\psi_{-}^{i}
+2\sqrt{-1}{\bar \psi}_{+}^{\bar i}\pa_{-}\psi_{+}^{i}
-4{\bar \phi}^{\bar i}\pa_{-}\pa_{+}\phi^{i}+F^{i}{\bar F}^{\bar i})\nonumber\\
&&-\frac{\pa W}{\pa \phi^{i}}F^{i}
-\frac{\pa {\bar W}}{\pa {\bar \phi}^{\bar i}} {\bar F}^{\bar i}
+\frac{\pa^{2}W}{\pa\phi^{i}\pa\phi^{j}}\psi_{+}^{i}\psi_{-}^{j}
+\frac{\pa^{2}{\bar W}} {\pa {\bar \phi}^{\bar i}\pa {\bar \phi}^{\bar j}}
{\bar \psi}_{-}^{\bar i}{\bar \psi}_{+}^{\bar j}.
\ENN
See Appendix for the convention of the $N=2$ superformalism .
 The equations of motion for the Landau-Ginzburg model can be written
 by superfields as
\EE
2{\bar D}_{+}{\bar D}_{-}{\bar \Phi}^{\bar i}=\frac{\pa W}{\pa \Phi^{i}}
\EN
In terms of the component fields we have
\EE
4F^{i}-\frac{\pa {\bar W}}{\pa {\bar \phi}^{\bar i}}=0,\ \ \
4{\bar F}^{\bar i}-\frac{\pa W}{\pa \phi^{i}}=0 \nonumber \\
\EN
\EE
8\sqrt{-1}\pa_{+}\psi_{-}^{i}+\pa_{\bar i}\pa_{\bar j}{\bar W}
{\bar \psi}_{+}^{\bar j}=0,\ \
-8\sqrt{-1}\pa_{+}{\bar \psi}_{-}^{\bar i}+\pa_{i}\pa_{j}W\psi_{+}^{j}=0
\EN
\EEN
&&8\sqrt{-1}\pa_{-}\psi_{+}^{i}-\pa_{\bar i}\pa_{\bar j}{\bar W}
{\bar \psi}_{-}^{\bar j}=0,\ \
-8\sqrt{-1}\pa_{-}{\bar \psi}_{+}^{\bar i}-\pa_{i}\pa_{j}W\psi_{-}^{j}=0 \\
&&-16\pa_{-}\pa_{+}\phi^{i}-\frac{1}{4}\pa_{i}W\pa_{\bar i}\pa_{\bar j}{\bar W}
+\pa_{\bar i}\pa_{\bar j}\pa_{\bar k}{\bar W}{\bar \psi}_{-}^{\bar j}
{\bar \psi}_{+}^{\bar k}=0 \\
&&-16\pa_{-}\pa_{+}{\bar \phi}^{\bar i}-\frac{1}{4}\pa_{\bar j}{\bar W}
\pa_{i}\pa_{j}W+\pa_{i}\pa_{j}\pa_{k}W\psi_{+}^{j}\psi_{-}^{k}=0
\ENN
The canonical energy-momentum tensor of the Landau-Ginzburg model reads
\EEN
T_{++}&=& \pa_{+}\phi^{i}\pa_{+}{\bar \phi}^{\bar i}
-\frac{\sqrt{-1}}{4}(\pa_{+}{\bar \psi}_{+}^{\bar i}\psi_{+}^{i}
-{\bar \psi}_{+}^{\bar i}\pa_{+}\psi_{+}^{i}) \nonumber \\
T_{+-} &=& \frac{1}{32}(\pa_{\bar i}\pa_{\bar j}{\bar W}
{\bar \psi}_{+}^{\bar i}
{\bar \psi}_{-}^{\bar j}+\pa_{i}\pa_{j}W\psi_{-}^{i}\psi_{+}^{j})
+\frac{1}{64}\pa_{i}W\pa_{\bar i}{\bar W} \nonumber \\
T_{--}&=& \pa_{-}\phi^{i}\pa_{-}{\bar \phi}^{\bar i}
-\frac{\sqrt{-1}}{4}(\pa_{-}{\bar \psi}_{-}^{\bar i}\psi_{-}^{i}
-{\bar \psi}_{-}^{\bar i}\pa_{-}\psi_{-}^{i}).\label{untwist}
\ENN
%We finally give the super transformations of component fields using the
%equation of motion of $F$ field.
%\EEN
%\delta \phi^{i}=\sqrt{2}(\xi^{-}\psi_{-}^{i}+\xi^{+}\psi_{+}^{i}),
%\delta{\bar \phi}^{\bar i}=-\sqrt{2}({\bar \xi}^{-}{\bar \psi}_{-}^{\bar i}
%+{\bar \xi}^{+}{\bar \psi}_{+}^{\bar i}) \\
%\delta\psi_{-}^{i}=\sqrt{2}(-\frac{1}{4}\xi^{+}\pa_{\bar i}{\bar W}
%-2\sqrt{-1}{\bar \xi}^{-}\pa_{-}\phi^{i}),
%\delta{\bar \psi}_{-}^{\bar i}=\sqrt{2}(-\frac{1}{4} {\bar \xi}^{+}\pa_{i}W
%+2\sqrt{-1}\xi^{-}\pa_{-}{\bar \phi}^{\bar i}) \\
%\delta\psi_{+}^{i}=\sqrt{2}(\frac{1}{4}\xi^{-}\pa_{\bar i}{\bar W }
%-2\sqrt{-1}{\bar \xi}^{+}\pa_{+}\phi^{i}),
%\delta{\bar \psi}^{\bar i}=\sqrt{2}(\frac{1}{4} {\bar \xi}^{-}\pa_{i}W
%+2\sqrt{-1}\xi^{+}{\bar \phi}^{\bar i})
%\ENN
%
\subsection{Topological twist  of Landau-Ginzburg models}
Let us consider a field theory which has an energy-momentun tensor
$(T_{--},T_{+-},T_{++})$ and a conserved current $(J_{-},J_{+})$.
Then one can define a new energy-momentum tensor as
\EEN
{\tilde T}_{--} & = & T_{--}-\frac{1}{2}\pa_{-}J_{-} \nonumber \\
{\tilde T}_{+-} & = & T_{+-}-\frac{1}{2}\pa_{-}J_{+}
               =  T_{+-}+\frac{1}{2}\pa_{+}J_{-}  \nonumber \\
{\tilde T}_{++}& =&  T_{++}+\frac{1}{2}\pa_{+}J_{+} \label{twisting}
\ENN
It is clear that the possibility of twist procedure depends upon
if the current is conserved.
We thus first look for conserved currents in the Landau-Ginzburg
model to consider its twists.
Two $U(1)$ transformations , which are essentially the rotations of
the super co-ordinates $(\th^{\pm},\thb^{\pm})$ ,
can be defined ;
the axial $U(1)$ transformation,
\EE
(\th^{-},\th^{+})\rightarrow  e^{\sqrt{-1}\beta}(\th^{-},\th^{+})
     ,\ \ ({\bar \th}^{-},{\bar \th}^{+})\rightarrow
e^{-\sqrt{-1}\beta}({\bar \th}^{-},{\bar \th}^{+})
\EN
and the vector $U(1)$ transformation,
\EE
(\th^{-},\th^{+})\rightarrow
(e^{-\sqrt{-1}\beta}\th^{-},e^{\sqrt{-1}\beta}\th^{+}) ,\ \
({\bar \th}^{-},{\bar \th}^{+})\rightarrow
(e^{\sqrt{-1}\beta}{\bar \th}^{-},e^{-\sqrt{-1}\beta}{\bar \th}^{+}).
\EN
It is seen that the vector $U(1)$ symmetry is conserved but the
axial $U(1)$ symmetry is violated by the superpotential term,
\EE
\frac{1}{2}d\th^{-}d\th^{+}\rightarrow e^{-2\sqrt{-1}\beta}
\frac{1}{2}d\th^{-}d\th^{+}.
\EN
The conserved current of  the vector $U(1)$ symmetry is given by
\EE
(J_{-}^{\rm vect},J_{+}^{\rm vect})
=(-\frac{\sqrt{-1}}{2}\psi_{-}^{i}{\bar \psi}_{-}^{\bar i},
-\frac{\sqrt{-1}}{2} {\bar \psi}_{+}^{\bar i}\psi_{+}^{i})\label{vector}.
\EN
Using the vector $U(1)$ current (\ref{vector}) we  twist the
energy-momentum tensor (\ref{untwist}) according to  (\ref{twisting}).
Then two of four supercharges ${\bar Q}_{\pm}$ turn into Lorentz scalars
 with erspect to  the twisted energy-momentum operator (\ref{twisting})
and we can define the BRS operator $Q_{BRS}$ as
\EE
Q_{BRS}={\bar Q}_{+}+{\bar Q}_{-}.
\EN
Now the twisted system has the following topological formula,
\EEN
{\tilde T}_{++}&= &\{Q_{BRS},\frac{\sqrt{-1}}{2\sqrt{2}}
\pa_{+}{\bar \phi}^{\bar i}\psi_{-}^{i}\}\nonumber \\
{\tilde T_{+-}}&= &\{Q_{BRS},-\frac{1}{16\sqrt{2}}\pa_{\bar i}{\bar W}
{\bar \phi}_{-}^{\bar i}\}  \nonumber   \\
{\tilde T}_{--}&=& \{Q_{BRS},\frac{\sqrt{-1}}{2\sqrt{2}}
\pa_{-}{\bar \phi}^{\bar i}\psi_{-}^{i}\}.
\ENN
This is the Vafa's topological Landau-Ginzburg model \cite{Va1}.
BRS local observables consist of the lowest component of chiral super fields
 which generate the chiral ring ${\cal R}$ \cite{LeVaWa}
\EE
{\cal R}\cong {\bf C}[\phi_{1},\phi_{2},..,\phi_{n}]/
(\frac{\pa W}{\pa \phi_{1}},...,\frac{\pa W}{\pa \phi_{n}})
     \label{chiral ring}
\EN
\subsection{Half-twisted Landau-Ginzburg models}
We now show that if one twists the energy-momentum tensor (\ref{untwist})
by the right-moving chiral $U(1)$ current then the remaining model has
left-moving conformal symmetry.
The right chiral $U(1)$ rotation is defined by
\EE
(\th^{-},{\bar \th}^{-})\rightarrow (\th^{-},{\bar \th}^{-}),\ \
(\th^{+},{\bar \th}^{+})\rightarrow
(e^{\sqrt{-1}\beta}\th^{+},e^{-\sqrt{-1}\beta}{\bar \th}^{+}).
\EN
Similarly to  the axial $U(1)$ rotation, the superpotential term violates
this symmetry .
However it is recovered if one can simaltaneously transform
the chiral superfields so that
\EE
W(\Phi^{i})\rightarrow W(e^{\sqrt{-1}\alpha_{i}\beta}\Phi^{i})
=e^{\sqrt{-1}\beta}W(\Phi^{i}),  \label{homo}
\EN
namely if $W$ is quasi-homogeneous. Here the weight $\alpha_{i}$ is called
the quasi-homogeneous degree of $\Phi^{i}$ \cite{VaWa}.
When the superpotential is quasi-homogeneous, therefore, we have
the conserved chiral $U(1)$ current
\EEN
J^{\rm R}_{+}&=& \frac{\alpha_{i}}{2} (-2{\bar \phi}^{\bar i}\pa_{+}\phi^{i}
+\sqrt{-1}{\bar \psi}_{+}^{\bar i}\psi_{+}^{i})-\frac{\sqrt{-1}}{2}
{\bar \psi}_{+}^{\bar i}\psi_{+}^{i} \nonumber \\
J^{\rm R}_{-}&=& \frac{\alpha_{i}}{2}(2\phi^{i}\pa_{-}{\bar \phi}^{\bar i}
+\sqrt{-1}{\bar \psi}_{-}^{\bar i}\psi_{-}^{i}). \label{right}
\ENN
Now let us  twist the energy-momentum tensor (\ref{untwist})
with the right moving chiral $U(1)$ current(\ref{right})
according to the  formula (\ref{twisting}). Define
\EE
Q_{BRS}={\bar Q}_{+}
\EN
as a BRS operator,
then we have the following energy-momentum operator
\EEN
T_{++}^{ht} &=& \frac{\sqrt{-1}}{2\sqrt{2}}
\{Q_{BRS},\pa_{+}{\bar \phi}^{\bar i}\psi_{+}^{i}
-\frac{1}{2}\alpha_{i}\pa_{+}({\bar \phi}^{\bar i}\psi_{+}^{i})\}\nonumber  \\
T_{+-}^{ht} &=& \{Q_{BRS},-\frac{\sqrt{-1}}{16\sqrt{2}}
{\bar \pa}_{\bar i}{\bar W}{\bar \psi}_{-}^{\bar i}
+\frac{\sqrt{-1}}{4\sqrt{2}}\alpha_{i}\pa_{-}({\bar \phi}^{\bar i}
\psi_{+}^{i})\}  \nonumber \\
T_{--}^{ht} &=& \pa_{-}\phi^{i}\pa_{-}{\bar \phi}^{\bar i}-\frac{\sqrt{-1}}
{4}(\pa_{-}{\bar \psi}^{\bar i}\psi_{-}^{i}-
{\bar \psi}_{-}^{i}\pa_{-}\psi_{-}^{i})\nonumber \\
&&-\frac{\alpha_{i}}{4}\pa_{-}(2\phi^{i}\pa {\bar \phi}^{\bar i}
+\sqrt{-1}{\bar \psi}_{-}^{\bar i}\psi_{-}^{i})  \nonumber \\
0 &=& \{Q_{BRS},T_{--}^{ht}\} ,\ \pa_{+}T_{--}^{ht}=\{Q_{BRS},*\}
\ENN
Only a (- -) component of $T^{ht}$ is an observable in the sense of
BRS cohomology . Thus in the half-twisted model conformal invariance
is realized in the left moving sector at the level of the $Q_{BRS}$
cohomology \cite{W1} \cite{W2}.
It can be checked  that $T_{--}^{ht}$ generates the Virasoro algebra with
${\hat c}=\sum_{i}(1-2\alpha_{i})$ .
There are infinite number of the  observables of the half-twisted
Landau-Ginzburg model and their spectra are represented by the elliptic
genus.  (See \cite{W2} \cite{DiYa} \cite{KaYaYa}  for the computations
of the elliptic genera of the various Landau-Ginzburg models.)
We conclude  that making use of the half-twist operation one can study
the $N=2$ superconformal algebra realized at the infra-red fixed point of
the Landau-Ginzburg model with a quasi-homogeneous superpotential $W$ .
In the following we search for $N=2$ super $W$ currents
in multi-component Landau-Ginzburg models by constructing local observables
as commutants of the  BRS operator $Q_{BRS}$ of half-twisted model.

\section{The $D_{2n+2}$-type minimal model}
\subsection{General strategy}
Our approach of constructing the  $N=2$ super $W$ algebra is
as simple as that for the  super Toda model \cite{NoMo}.
What we should do is to find operators expressed as  differential polynomials
of chiral superfields which are in the $Q_{BRS}$ cohomology class.
According to the equivalence relation \cite{W2},
${\displaystyle {\bar Q}_{+}=e^{-4\sqrt{-1}\thb^{+}\th^{+}\pa_{+}}{\bar D}_{+}
e^{4\sqrt{-1}\thb^{+}\th^{+}\pa_{+}}}$
we  shall consider cohomology classes of ${\bar D}_{+}$ for convenience.
To find a  BRS cohomology of spin {\it m}  and charge 0
 we take as the candidates all the monomials
\EE
\prod (\Phi^{i})^{a(i)}(\DM\Phi^{i})^{b(i)}(\DMB{\bar \Phi}^{\bar i})^{c(i)}
(\DM\DMB\Phi^{\ib})^{d(i)}
\EN
such that $\sum \alpha_{i}(a(i)+b(i)-c(i)-d(i))=0,
\ \sum (c(i)+d(i))=m \
\ \sum (b(i)+d(i))=m $.
Similarly BRS exact operators are made by operating ${\bar D}_{+}$
to monomials such that
$
\sum \alpha_{i}(a(i)+b(i)-c(i)-d(i))=-1 ,
\ \ \sum (c(i)+d(i))=m+1,\ \
\sum (d(i)+b(i))=m .
$
In this paper we only  deal with the  classical aspect of the algebra  and
 possible quantum corrections to observables are neglected.
Before analyzing the multi-variable case we briefly review the 1-variable case.
The model is the A-type minimal model with the superpotential
\EE
W(X)=\frac{1}{k+2}X^{k+2}.
\EN
The  equation of motion is
\EE
2\DPB\DMB {\bar X}=X^{k+1}. \label{exact}
\EN
The energy-momentum superfield obtained by  Witten is
\EE
J=\frac{k+1}{2(k+2)}\DM X\DMB \XB-\frac{1}{2(k+2)}X\DM\DMB\XB,\  \ \DPB J=0.
\EN
We easily see that $J^{k+1}$ is a BRS exact operator since
\EE
(-2(k+1)J)^{k+1}=\DPB(2\DMB\XB (\DM\DMB\XB)^{k+1}) \label{null}.
\EN
%Is should be noted that (\ref{null}) looks like (\ref{exact}), which
%would imply a existence of a simple transformation law from chiral super
%fields to $N=2$ super $W$ operators.
%This correspondence is extended to other multi-variable cases below.
It is interesting to notice that in (\ref{exact}) the superfield $X$
raised to the power $k+1$ is BRS exact while in (\ref{null})
the energy-momentum superfield $J$ raised to the same power $k+1$ is also
BRS exact. In this sense we observe simple correspondence between
the chiral superfield and the $N=2$ $W$ current.
This kind of correspondence is not peculiar to the A-type model,
but can be extended to  other multi-variable cases as we will see below.

\subsection{The $D_{2n+2}$ model}
The superpotential and the energy-momentum tensor for the $D_{2n+2}$-type
minimal  model is given by
\EEN
W(X,Y) &=& \frac{1}{2n+1}X^{2n+1}+XY^{2} \nonumber  \\
2(2n+1)J &=& 2n\DM X\DMB\XB-X\DM\DMB\XB
+(n+1)\DM Y\DMB\YB-nY\DM\DMB\YB
\ENN
%According to the conjecture of \cite{VaWa} $N=2$ super conformal algebra
%of the model at the infra-red fixed point has
%central charge  and belongs to $N=2$
% minimal series.
This model at the infra-red fixed point has central charge
${\hat c}={\displaystyle \frac{2n}{2n+1}}$  \cite{VaWa}
and it has a realization as
the  level 1 supercoset $SO(2n+2)/SO(2n)\times SO(2)$ model \cite{KaSu}
\cite{EgKaMiYa}.
{}From the form of the modular invariant partition function one  expects that
in this model $N=2$ algebra is extended by the  operator of spin {\it n}
\cite{FeLeMaWa}. We construct directly this operator for $n=1,2,3,4$ below.
These  are all parity odd under the symmetry $Y \rightarrow -Y$.
\\
$\underline{ n=1 {\rm case} }$ \\
%Super potential, equations of motion, and energy-momentum operator are
%\EEN
%W(X,Y) &=& \frac{1}{3}X^{3}+XY^{2} \nonumber \\
%2\DPB\DMB\XB &=& X^{2}+Y^{2},\ \ 2\DPB\DMB\YB=2XY \nonumber \\
%6J &=& 2\DM X\DMB \XB-X\DM\DMB\XB +2\DM Y\DMB\YB-Y\DM\DMB\YB
%\ENN
%In fact this model  trivially  decouples by the change of variables
%$X=\frac{1}{2}(N+M),\ Y=\frac{1}{2}(N-M)$.
One has the following BRS cohomology class of spin 1 with odd
${\bf Z}_{2}$ parity,
\EE
J_{A}=\frac{1}{6}(2DX\DB \YB +2DY\DB\XB-XD\DB\YB-YD\DB\XB)
\EN
%with $sl(2\mid 1)$ \cite{NoMo}.
It is readily seen that $J and J_A$ generate a ring with respect to the
classical product
\EEN
(3J)(3J_{A}) &=& \DPB\left [ \frac{1}{4}(D\DB\XB)^{2}\DB\YB
+\frac{1}{2}\DB\XB D\DB\XB D\DB\YB
+\frac{1}{4}\DB\YB(D\DB\YB)^{2}\right ]\nonumber \\
(3J)^{2}+(3J_{A})^{2} &=& \DPB\left [ \frac{1}{2}\DB\XB(D\DB\XB)^{2}
+D\DB\XB\DB\YB D\DB\YB
+\frac{1}{2}\DB\XB(D\DB\YB)^{2} \right ].
\ENN
Notice that this is isomorphic to the chiral ring
\EEN
2\DPB\DMB\XB &=& \frac{\pa W}{\pa X}= X^{2}+Y^{2}, \nonumber \\
2\DPB\DMB\YB &=& \frac{\pa W}{\pa Y}= 2XY.
\ENN
Before going to the $n=2$ case, we describe the quantum version
of  the extended algebra.
Let us  renormalize the $W$ supercurrents as
\EE
T=-\frac{4\pi}{8}J, \ T_{A}=-\frac{4\pi}{8}J_{A}
\EN
(see Appendix), then we evaluate the operator product expansions
\EEN
T(Z_{1})T_{A}(Z_{2}) &=& \left[\frac{\th_{12}\thb_{12}}{Z_{12}^{2}}
+\frac{\sqrt{-1}}{4}\frac{\th_{12}}{Z_{12}}D+\frac{\sqrt{-1}}{4}
\frac{\thb_{12}}{Z_{12}}\DB
+\frac{\th_{12}\thb_{12}}{Z_{12}}\pa\right]T_{A}(Z_{2})\nonumber \\
T(Z_{1})T(Z_{2}) &=& -\frac{1}{16} \frac{2}{3}+
\left [ \frac{\th_{12}\thb_{12}}{Z_{12}^{2}}
+\frac{\sqrt{-1}}{4}\frac{\th_{12}}{Z_{12}}D
+\frac{\sqrt{-1}}{4}\frac{\thb_{12}}{Z_{12}}\DB
+\frac{\th_{12}\thb_{12}}{Z_{12}}\pa \right ]T(Z_{2})      \nonumber \\
T_{A}(Z_{1})T_{A}(Z_{2}) &=& -\frac{1}{16} \frac{2}{3}+
\left [ \frac{\th_{12}\thb_{12}}{Z_{12}^{2}}
+\frac{\sqrt{-1}}{4}\frac{\th_{12}}{Z_{12}}D
+\frac{\sqrt{-1}}{4}\frac{\thb_{12}}{Z_{12}}\DB
+\frac{\th_{12}\thb_{12}}{Z_{12}}\pa \right ]T(Z_{2}). \nonumber
\ENN
This algebra coincides with that of the $N=2$ super Toda field theory
associated  with the Lie superalgebra $sl(2\mid 1)$ \cite{NoMo}.
\\
$\underline{n=2 {\rm case} }$ \\
This model has ${\hat c}={\displaystyle \frac{4}{5}}$ and can be identified
with the level 2 $CP_{2}$ model.
 In this case we search for a spin 2 BRS cohomology
element in the parity odd sector.
There exist  two BRS closed operators, one of which is BRS exact and
the other is not.
The BRS exact operator is
\EE
-2X^{2}DY\DB\YB D\DB\YB+2XY\DB\YB DXD\DB\YB
+X^{2}(D\DB\YB)^{2}
= \DPB(X\DB\YB(D\DB\YB)^{2}), \nonumber
\EN
and the BRS observable $W^{(2)}$ is
\EEN
&&W^{(2)} = 2DY\DB\XB D\DB\XB-\frac{1}{2}Y(D\DB\XB)^{2}
-X^{2}Y(D\DB\YB)^{2}
+3X^{2}DY\DB\YB D\DB\YB  \nonumber  \\
&&-X^{3}D\DB\XB D\DB\YB+4X^{2}DX\DB\XB D\DB\YB
+3X^{2}D XD\DB\XB\DB\YB.
\ENN
 We find  that $10J$ and $2W^{(2)}$ form a ring under classical product
\EEN
(10J)(2W^{(2)}) &=& \DPB \big[\cdots \big] \nonumber \\
(10J)^{4}+(2W^{(2)})^{2} &=& \DPB \big[\cdots \big].\label{n2}
\ENN
Notice again that this is isomorphic to the chiral ring,
\EEN
2\DPB\DMB\XB &=& \frac{\pa W}{\pa X} = X^{4}+Y^{2}\nonumber \\
2\DPB\DMB\YB &=& \frac{\pa W}{\pa Y} = 2XY .
\ENN
The detailed description of the right hand side of (\ref{n2}) is given in
Appendix.
\\
$\underline{n=3  {\rm case} }$  \\
In this model we obtain the  supercurrent of spin 3,
\EEN
&&W^{(3)} = \frac{1}{3}Y(D\DB \XB)^{3}-2DY\DB\XB(D\DB\XB)^{2}\nonumber \\
&&+X^{5}D\DB\YB(D\DB\XB)^{2}-4X^{4}DX\DB\YB(D\DB\XB)^{2}
-12X^{4}DX\DB\XB D\DB\XB D\DB\YB  \nonumber  \\
&&-6X^{4}DY\DB\XB(D\DB\YB)^{2}-8X^{4}DYD\DB\YB\DB\YB D\DB\XB
+3X^{4}YD\DB\XB(D\DB\YB)^{2} \nonumber  \\
&&-12X^{8}DX\DB\YB(D\DB\YB)^{2}+X^{9}(D\DB\YB)^{3}.
\ENN
Moreover there exist four BRS exact currents of spin 3
in the parity odd sector,
\EEN
&&\DPB(X^{2}Y\DB\YB(D\DB\YB)^{3}),\ \ \DPB(X^{3}\DB\XB(D\DB\YB)^{3})
 \nonumber  \\
&&\DPB(X^{2}DX\DB\XB\DB\YB(D\DB\YB)^{2}), \ \
\DPB(X^{3}\DB\YB D\DB\XB(D\DB\YB)^{2}). \nonumber   \\
\ENN
These  super $W$ currents generate  the ring  under the classical product,
\EEN
(14J)(3W^{(3)}) &=& \DPB \big[\cdots \big]\nonumber \\
(14J)^{6}+(3W^{(3)})^{2} &=& \DPB \big[\cdots \big],\label{n3}
\ENN
 which is isomorphic to the chiral ring
\EEN
2\DPB\DMB\XB &=& \frac{\pa W}{\pa X} = X^{6}+Y^{2}, \nonumber \\
2\DPB\DMB\YB &=& \frac{\pa W}{\pa Y} = 2XY.
\ENN
We represent the right hand side of (\ref{n3}) explicitly in Appendix. \\
$\underline{n=4 {\rm case} }$\\
We find here that there exists a unique spin 4 BRS cohomology element
with the odd parity,
\EEN
W^{(4)} &=&
\frac{1}{4}Y(D\DB\XB)^{4}-2DY\DB\XB(D\DB\XB)^{3}\nonumber  \\
&&-5X^{6}DX\DB\YB(D\DB\XB)^{3}
-24X^{6}DXD\DB\YB\DB\XB(D\DB\XB)^{2} \nonumber  \\
&&+X^{7}D\DB\YB(D\DB\XB)^{3}
-48X^{5}DXDY\DB\XB\DB\YB D\DB\XB D\DB\YB \nonumber  \\
&&+12X^{5}YDX\DB\YB D\DB\YB(D\DB\XB)^{2}
-3X^{6}DY\DB\YB D\DB\YB(D\DB\XB)^{2}
\ENN
In addition to this, other 9 BRS exact terms of spin 4
with odd parity  exist though we don't describe here.
%We suspect that $18J$ and $4W^{(4)}$ generate a ring
%isomorphic to the chiral ring .
%
%From the description of a few examples one may expect that the super current
%of spin n $W^{(n)}$ would be given in the general form .
Now  we would like to  conjecture that the $D_{2n+2}$-type minimal model has a
spin {\it  n} parity odd current of the form,
\EEN
&&W^{(n)}=\frac{1}{n}Y(D\DB\XB)^{n}-2DY\DB\XB(D\DB\XB)^{n-1}
+X^{2n-1}(D\DB\XB)^{n-1}D\DB\YB  \\
&&-(n+1)X^{2n-2}DX(D\DB\XB)^{n-1}\DB\YB
 -2n(n-1)X^{2n-2}DX\DB\XB(D\DB\XB)^{n-2}D\DB\YB +\cdots \nonumber
\ENN
where the remainig terms in the ellipses can be systematically  evaluated
for each $n$.
We also  conjecture  that under the classical product the supercurrents
 $(2(2n+1)J,nW^{(n)})$ generate a ring graded by $L_{0}$ which is isomorphic
to the chiral ring ${\cal R}$ of the model generated by $(X,Y)$, i.e.,
\EEN
(2(2n+1)J)(nW^{(n)}) &=& \DPB \big[\cdots \big] \nonumber \\
(2(2n+1)J)^{2n}+(nW^{(n)})^{2} &=& \DPB\big[\cdots\big].
\ENN
It would be interesting if one can find the Miura transformation
$(X,Y)\rightarrow (J,W^{(n)})$. For the case of  $CP_n$ coset model
we have such a transformation as will be explained in the next section.

\section{The $CP_{n}$ coset model}
\subsection{General aspect}
For the $CP_{n}$ coset model , Ito \cite{It1} identified the
$N=2$ super $W$  symmetry
of the model with that of super Toda field theory \cite{NoMo}
\cite{KoMoNo} \cite{EHo}.
 This super $W$ algebra has the  generators of spin $\{1,2,3,...,n\}$.
In the super Toda approach, these generators are given by the
Miura transformation \cite{EHo}	, which originates from the fact that
the super Toda field equation  can be written in the form of super flat
curvature equation.
In the following we will give  simple formulas for the BRS invariant
supercurrents of spin $\{1,2,3,...,n\}$ in terms of the chiral superfields.
(The relation between these two approaches  is  yet to be clarified.)\\
The chiral ring of $CP_{n}^{(k)}$ model is isomorphic to the cohomology ring
of the coset space $SU(n+k+1)/SU(n+1)\times SU(k)\times U(1)$ \cite{LeVaWa}.
An important property is the level-rank duality
\EE
CP_{n}^{(k)} \cong CP_{k}^{(n)} ,
\  {\hat c}=\frac{kn}{n+k+1}.\label{level}
\EN
The chiral superfields and the superpotential of $CP_{n}^{(k)}$ model are
given as follows \cite{Ge2},
%\EEN
%\Phi^{1} &=& \sum_{i} a(i)=a(1)+a(2)+\cdots +a(n) \nonumber \\
%\Phi^{2} &=& \sum_{i_{1}\lt i_{2}}a(i_{1})a(i_{2}) \nonumber \\
\EE
1+t\Phi^{1}+t^{2}\Phi^{2}+\cdots t^{n}\Phi^{n}=
(1+ta(1))(1+ta(2))\cdots (1+ta(n)) \label{split}
\EN
%\EE
%\Phi^{l}=\sum_{a(i_{1})\lt a(i_{2}) \lt \cdots a(i_{l})}
%a(i_{1})a(i_{2})\cdots a(i_{l}) , \ \ 1\geq l \le n,
%\EN
%
\EEN
\Phi^{1} &=& a(1)+a(2)+ \cdots +a(n) \nonumber  \\
\Phi^{2} &=& a(1)a(2)+a(1)a(3)+\cdots +a(n-1)a(n) \nonumber \\
\vdots   & & \vdots                 \nonumber  \\
\Phi^{n} &=& a(1)a(2)\cdots a(n)
\ENN
\EE
W_{(k+n+1)}=\frac{1}{(n+k+1)}\sum_{l=1}^{n}a(l)^{n+k+1}. \label{potential}
\EN
The point is that the right hand side of (\ref{potential}) is a
completely symmetric polynomial of $a(1),a(2),\ldots a(n)$ so that it can be
expanded as a polynomial of $\Phi^{1},\Phi^{2},\ldots \Phi^{n}$.
Our idea to construct the BRS closed supercurrents of spin ${1,2,\ldots n}$
is very simple.
First one  writes the equation of motion  in terms of  chiral superfields
$\{a(1),a(2),\ldots a(n)\}$ as
\EE
2\DPB\DMB {\bar b}(\ib) = a(i)^{k+n},\ \ 1 \leq i \leq n \ \ ,
{\bar b}(\ib) = \frac{\pa K}{\pa a(i)} ,
\ \ K=\sum \Phi^{i}{\bar \Phi}^{\ib} \label{aaa}
\EN
It is easily seen that (\ref{aaa}) is equivalent to the original equations
of motion of $\Phi^{i}$s.
Next we observe that there are  {\it n} BRS invariant spin 1  currents
\EE
J(i)=(k+n)Da(i)\DB {\bar b}(\ib)-a(i)D\DB{\bar b}(\ib),
\ \ 1 \leq i \leq n
\EN
\EE
\DPB J(i)=0 , \ \ \ 1 \leq i \leq n
\EN
 If one combine these {\it n} currents in a completely symmetric manner
 in  ${\it i}$ ,
\EE
1+tW^{(1)}+t^{2}W^{(2)}+t^{3}W^{(3)}+\cdots t^{n}W^{(n)}
= (1+tJ(1))(1+tJ(2))\cdots (1+tJ(n)) \label{split2}
\EN
 the resulting spin ${\it i}$ current $W^{(i)}$ can be rewritten by
the original variables $\Phi$. Then $W^{(i)}$ s are BRS invariant by
construction.
In particular $W^{(1)}$ coincides with the energy-momentum operator
\EE
W^{(1)}=2(n+k+1)J=\sum_{i=1}^{n}\bigl[(n+k+1)D\Phi^{i}\DB{\bar \Phi}^{\ib}
-iD(\Phi^{i}\DB{\bar \Phi}^{\ib})\bigl],
\EN
The remaining higher spin currents $W^{(i)}$ should give an extention of
 the $N=2$ superconformal algebra under the Poisson bracket.
We shall check this for the $CP_2$ and $CP_3$ models later.
Finally we remark that the transformation (\ref{split2}) is the analogue
of the Miura transformation.
\subsection {The $CP_{2}$ model}
The superpotential and chiral superfields  are given by
\EE
W_{k+3}=\frac{1}{k+3}(a(1)^{k+3}+a(2)^{k+3}),\ \ X = a(1)+a(2), \ Y = a(1)a(2).
\EN
The kinetic term (K$\ddot{\rm a}$hler potential) is written as
\EE
K=X\XB+Y\YB
=(a(1)+a(2))({\bar a(1)}+{\bar a(2)})+a(1)a(2){\bar a(1)}{\bar a(2)},
\EN
so that
\EE
{\bar b(1)} = \frac{\pa K}{\pa a(1)}=\XB +a(2)\YB ,\ \
{\bar b(2)} = \frac{\pa K}{\pa a(2)}=\XB +a(1)\YB.
\EN
Two spin 1  currents are
\EEN
J(1) &=& (k+2)Da(1)\DB{\bar b(1)}-a(1)D\DB{\bar b(1)} \nonumber  \\
J(2) &=& (k+2)Da(2)\DB{\bar b(2)}-a(2)D\DB{\bar b(2)},
\ENN
from which one can construct BRS invariant supercurrent of spin 1 and 2,
\EE
W^{(1)}=2(k+3)J=(k+2)DX\DB\XB +(k+1)DY\DB\YB -XD\DB\XB -2YD\DB\YB
\EN
\EEN
W^{(2)}&=&-(k+2)^{2}DXDY\DB\XB\DB\YB
-(k+2)YDX\DB\XB D\DB\YB
-(k+2)XDY\DB D\DB\XB      \nonumber    \\
&& +(k+3)YDXD\DB\XB\DB\YB
 +XYD\DB\XB D\DB\YB  -(k+2)DY\DB\XB D\DB\YB \nonumber    \\
&&+Y(D\DB\XB)^{2}
-(k+1)YDY\DB\YB D\DB\YB
+Y^{2}(D\DB\YB)^{2}
\ENN
As in the $D_{2n+2}$-type minimal model treated in the
previous section, one may suspect that $W^{(1)}$ and
$W^{(2)}$ generate a ring isomorphic to the chiral ring with respect to
 the classical product.
Indeed this is true as we show. To this purpose we describe the ideal
of the chiral ring \cite{Ge2}.
The generators of thre ideal of the chiral ring of the level $k$
model can easily described by the variables $(a(1),a(2))$ as
\EEN
\frac{\pa W_{k+3}}{\pa X} &=&
a(1)^{k+2}+a(1)^{k+1}a(2)+a(1)^{k}a(2)^{2}+\cdots +a(2)^{k+2}\nonumber \\
-\frac{\pa W_{k+3}}{\pa Y} &=&
a(1)^{k+1}+a(1)^{k}a(2)+a(1)^{k-1}a(2)^{2}+\cdots +a(2)^{k+1}. \label{idee}
\ENN
It is easily found that
\EEN
(-2(k+3)J(1))^{k+2} &=&
2\DPB [\DB {\bar b}(1)(D\DB{\bar b}(1))^{k+2}] \nonumber  \\
(-2(k+3)J(2))^{k+2} &=&
2\DPB [\DB {\bar b}(2)(D\DB{\bar b}(2))^{k+2}].  \label{big}
\ENN
However the $(1,2)-$ permutation invariant subideal of the ideal
$(J(1)^{k+2},J(2)^{k+2})$ is smaller than the ideal (\ref{idee})
 with $(a(1),a(2))$ replaced by $(J(1),J(2))$,
so that one need  a refinement of (\ref{big}).
The solution is given by the following equation
\EEN
&&\DB {\bar b}(1)(D\DB{\bar b}(1))^{k+2}
-\DB {\bar b}(2)(D\DB{\bar b}(2))^{k+2}\nonumber \\
&=&+2(k+3)(J(1)-J(2))M^{(k)}
-(a(1)-a(2))\DPB [\DB \XB\DB\YB(D\DB\YB)^{k+2}] \label{anti}
\ENN
 where $M^{(k)}$ is some superdifferential polynomial, i.e.,
an element of ${\bf C}[X,Y,\DB\XB,\DB\YB ;D]$.
{}From this  we have  the needed equation,
\EEN
2\DPB M^{(k)} &=& -\frac{(-2(k+3)J(1))^{k+2}-(-2(k+3)J(2))^{k+2}}
{-2(k+3)J(1)+2(k+3)J(2)} \nonumber  \\
&=& (-2(k+3))^{k+1}[J(1)^{k+1}+J(1)^{k}J(2)+\cdots +J(2)^{k+1}]. \label{ide}
\ENN
% , right hand side of which is presicely one of the generators of the ideal.
Now it is easy to see that  (\ref{ide}) combined with (\ref{big}) constitute
the ideal isomorphic to that of the chiral ring.
Thus we can say that $(W^{(1)},W^{(2)})$ generate a ring isomorphic to
the chiral ring under the classical product.
For the computation of $M^{(k)}$ see Appendix. \\
For example the generators of the ideal of the chiral ring for $k=1$
is given by
\EE
\frac{\pa W_{4}}{\pa X} = X^{3}-2XY ,\ \
\frac{\pa W_{4}}{\pa Y} = X^{2}-Y.
\EN
%Indeed  the above conjecture is true, as we show below, and
%the spin 2 current $W^{(2)}$ is BRS cohomologous to $(W^{(1)})^{2}$
%and $(W^{(1)})^{3}$ is cohomologous to 0 , which is consistent with the fact
%that the level 1 model is equivalent to the minimal model of level 2.
%Indeed we checked the BRS exactness of $(W^{(2)})-(W^{(1)})^{2}$
%by the direct computation.
Correspondingly,  the spin 2 current $W^{(2)}$ is BRS equivalent to
$(W^{(1)})^{2}$,
\EEN
&&(W^{(2)})-(W^{(1)})^{2}=2\DPB \bigl[(D\DB\XB)^{2}\DB\YB
+3\DB\XB D\DB\XB D\DB\YB \nonumber \\
&&+3DX\DB\XB\DB\YB D\DB\YB+3XD\DB\XB\DB\YB D\DB\YB
+3X\DB\XB(D\DB\YB)^{2}\nonumber \\
&&-3Y\DB\YB(D\DB\YB)^{2}+3X^{2}\DB\YB(D\DB\YB)^{2}\bigl]=2\DPB M^{(1)}
\ENN
and $(W^{(1)})^{3}$ is cohomologous to 0,
\EEN
&&(W^{(1)})^{3}=\DPB\bigl[72X^{2}DX\DB\XB\DB\YB(D\DB\YB)^{2}
-24X^{3}D\DB\XB\DB\YB(D\DB\YB)^{2} \nonumber \\
&&-16X^{2}Y\DB\YB(D\DB\YB)^{3}-24XDX\DB\XB D\DB\XB\DB\YB D\DB\YB \nonumber \\
&&-12X^{2}(D\DB\XB)^{2}\DB\YB D\DB\YB-24X^{2}\DB\XB D\DB\XB(D\DB\YB)^{2}
\nonumber \\
&&-72YDX\DB\XB\DB\YB(D\DB\YB)^{2}-48XDY\DB\XB\DB\YB(D\DB\YB)^{2}
\nonumber \\
&&+24XYD\DB\XB\DB\YB(D\DB\YB)^{2}-16XY\DB\XB(D\DB\YB)^{3} \nonumber \\
&&+16Y^{2}\DB\YB(D\DB\YB)^{3}-2\DB\XB(D\DB\XB)^{3} \nonumber \\
&&+12DX\DB\XB(D\DB\XB)^{2}\DB\YB-4X(D\DB\XB)^{3}\DB\YB \nonumber \\
&&-12X\DB\XB(D\DB\XB)^{2}D\DB\YB\bigl],
\ENN
this is consistent with the fact that the level 1 model is isomorphic to
the minimal model of level 2 (\ref{level}).
%
%
%
%
%
%
%
%
%
%
%
%
%
%
%
%
%]
%
%
%
%
%
%
%
%
Also in the general $CP_{n}^{(k)}$ model ,it seems that
the chiral ring conjecture
about  $\{W^{(1)},W^{(2)},\ldots ,W^{(n)}\}$
does not conflict with the level-rank duality (\ref{level} ).
\subsection{The $CP_{3}$ model}
The superpotential and the chiral superfields  are given by
\EEN
W_{k+4} &=& \frac{1}{(k+4)}(a(1)^{k+4}+a(2)^{k+4}+a(3)^{k+4})\nonumber  \\
X  &=& a(1)+a(2)+a(3) \nonumber \\
Y &=& a(1)a(2)+a(1)a(3)+a(2)a(3) \nonumber \\
Z &=& a(1)a(2)a(3).
\ENN
The spin 1 currents are
\EEN
J(1) &=& (k+3)Da(1)\DB{\bar b(1)}-a(1)D\DB{\bar b(1)} \nonumber  \\
J(2) &=& (k+3)Da(2)\DB{\bar b(2)}-a(2)D\DB{\bar b(2)}\nonumber \\
J(3) &=& (k+3)Da(3)\DB{\bar b(3)}-a(3)D\DB{\bar b(3)}\nonumber \\
{\bar b(1)}&=&\frac{\pa K}{\pa a(1)}
=\XB+(a(2)+a(3))\YB+a(2)a(3)\ZB \nonumber \\
{\bar b(2)}&=&\frac{\pa K}{\pa a(2)}
=\XB+(a(1)+a(3))\YB+a(1)a(3)\ZB \nonumber \\
{\bar b(3)}&=&\frac{\pa K}{\pa a(3)}
=\XB+(a(2)+a(1))\YB+a(2)a(1)\ZB
\ENN
We recover the energy-momentum operator as
\EEN
&&W^{(1)} = 2(k+4)J=J(1)+J(2)+J(3) \nonumber  \\
&=&(k+3)DX\DB\XB+(k+2)DY\DB\YB+(k+1)DZ\DB\ZB-XD\DB\XB-2YD\DB\YB-3ZD\DB\ZB.
\nonumber
\ENN
Computations of super $W$ currents $W^{(2)}=J(1)J(2)+J(1)J(3)+J(2)J(3)$,\\
$W^{(3)}=J(1)J(2)J(3)$ are a little complicated. The explicit results are
\EEN
&&W^{(2)}=-(k+3)^{2}DXDY\DB\XB\DB\YB
-(k+2)(k+3)DXDZ\DB\XB\DB\ZB \nonumber  \\
&&-(k+2)^{2}DYDZ\DB\YB\DB\ZB   \nonumber  \\
&&-(k+3)DY\DB\XB D\DB\XB-(k+3)YDX\DB\XB D\DB\YB
-(k+3)DZ\DB\XB D\DB\YB  \nonumber  \\
&&-2(k+3)ZDX\DB\XB D\DB\ZB-(k+3)XDY D\DB\XB\DB\YB
+(k+4)YDX D\DB\XB\DB\YB \nonumber  \\
&&-(k+2)DZ D\DB\XB\DB\YB-(k+2)YDY\DB\YB D\DB\YB
-(k+3)XDZ\DB\YB D\DB\YB \nonumber  \\
&&+(k+5)ZDX\DB\YB D\DB\YB-2(k+2)ZDY\DB\YB D\DB\ZB
-(k+2)XDZ D\DB\XB\DB\ZB \nonumber  \\
&&+(k+4)ZDX D\DB\XB\DB\ZB-2(k+2)YDZ D\DB\YB\DB\ZB
+(k+5)ZDY D\DB\YB\DB\ZB \nonumber  \\
&&-2(k+1)ZDZ\DB\ZB D\DB\ZB+Y(D\DB\XB)^{2}
+XYD\DB\XB D\DB\YB   \nonumber  \\
&&+3ZD\DB\XB D\DB\YB +2XZD\DB\XB D\DB\ZB
+Y^{2}(D\DB\YB)^{2}+XZ(D\DB\YB)^{2} \nonumber  \\
&&+2YZD\DB\YB D\DB\ZB+3Z^{2}(D\DB\ZB)^{2}
\ENN
\EEN
&&W^{(3)}=(k+3)^{3}DXDYDZ\DB\XB\DB\YB\DB\ZB \nonumber  \\
&&+(k+3)^{2}DXDZ\DB\XB D\DB\XB\DB\YB
+(k+3)^{2}DYDZ\DB\XB D\DB\XB\DB\ZB \nonumber  \\
&&+(k+3)^{2}(ZDXDY+XDYDZ-YDXDZ)D\DB\XB\DB\YB\DB\ZB \nonumber \\
&&+(k+3)^{2}(XDXDZ-DYDZ)\DB\XB\DB\YB D\DB\YB \nonumber  \\
&&+(k+3)^{2}(YDXDZ-ZDXDY)\DB\XB D\DB\YB\DB\ZB
+2(k+3)^{2}DXDY\DB\XB\DB\YB D\DB\ZB \nonumber  \\
&&+(k+3)^{2}ZDXDZ\DB\XB\DB\ZB D\DB\ZB
+(k+3)^{2}ZDYDZ\DB\YB\DB\ZB D\DB\ZB \nonumber  \\
&&+(k+3)^{2}(YDYDZ-ZDXDZ)\DB\YB D\DB\YB\DB\ZB \nonumber \\
&&+(k+3)DZ\DB\XB(D\DB\XB)^{2}
+(k+3)DXDZ\DB\XB D\DB\XB\DB\YB \nonumber  \\
&&+(k+3)(XDZ+ZDX)\DB\XB D\DB\XB D\DB\YB
+(k+3)ZDY\DB\XB D\DB\XB D\DB\ZB \nonumber \\
&&+(k+3)DYDZ\DB\XB\DB\YB D\DB\YB
+(k+3)XZDX\DB\XB(D\DB\YB)^{2} \nonumber  \\
&&-(k+3)ZDXDY\DB\XB D\DB\YB\DB\ZB
+(k+3)(YZDX+ZDZ)\DB\XB D\DB\YB D\DB\ZB \nonumber \\
&&-(k+3)ZDXDZ\DB\XB\DB\ZB D\DB\ZB
+(k+3)Z^{2}DX\DB\XB(D\DB\ZB)^{2} \nonumber \\
&&+(k+3)(XDZ-ZDX)(D\DB\XB)^{2}\DB\YB
+(k+3)(X^{2}DZ-XZDX+ZDY)D\DB\XB\DB\YB D\DB\YB \nonumber \\
&&+(k+3)ZDXDY D\DB\XB\DB\YB\DB\ZB
+(k+3)(XZDY-YZDX+ZDZ)D\DB\XB\DB\YB D\DB\ZB \nonumber  \\
&&+(k+3)(XY-Z)DZ\DB\YB(D\DB\YB)^{2}   \nonumber  \\
&&+(k+3)(YDYDZ+2XZDXDY-2XYDXDZ)\DB\YB D\DB\YB\DB\ZB \nonumber \\
&&+(k+3)(XZDZ+YZDY-Z^{2}DX)\DB\YB D\DB\YB D\DB\ZB
-2(k+3)ZDYDZ\DB\YB\DB\ZB D\DB\ZB \nonumber \\
&&+(k+3)Z^{2}DY\DB\YB(D\DB\YB)^{2}
+(k+3)(ZDXDY-YDXDZ)D\DB\XB\DB\YB\DB\ZB \nonumber \\
&&+(k+3)(YDZ-ZDY)(D\DB\XB)^{2}\DB\ZB
+(k+3)(XYDZ-XZDY+ZDZ)D\DB\XB D\DB\YB\DB\ZB \nonumber \\
&&+(k+3)(XZDZ-Z^{2}DX)D\DB\XB\DB\ZB D\DB\ZB
+(k+3)(Z^{2}DX+Y^{2}DZ-YZDY)(D\DB\YB)^{2}\DB\ZB \nonumber \\
&&+(k+3)(2YZDZ-Z^{2}DY)D\DB\YB\DB\ZB D\DB\ZB
+(k+3)Z^{2}DZ\DB\ZB(D\DB\ZB)^{2} \nonumber  \\
&&-Z(D\DB\XB)^{3}-2ZDX(D\DB\XB)^{2}\DB\YB-2XZ(D\DB\XB)^{2}D\DB\YB\nonumber \\
&&-ZDY(D\DB\XB)^{2}\DB\ZB-ZD(X^{2}+Y)D\DB\XB \DB\YB D\DB\YB \nonumber \\
&&+ZDXDYD\DB\XB\DB\YB\DB\ZB-Z(YDX+DZ)D\DB\XB\DB\YB D\DB\ZB \nonumber \\
&&-Z(X^{2}+Y)D\DB\XB(D\DB\YB)^{2}-Z(2DZ+XDY)D\DB\XB D\DB\YB\DB\ZB \nonumber \\
&&-Z(XY+3Z)D\DB\XB D\DB\YB D\DB\ZB-Z(ZDX+XDZ)D\DB\XB\DB\ZB D\DB\ZB\nonumber \\
&&-XZ^{2}D\DB\XB(D\DB\ZB)^{2}-Z(XDY+YDX-DZ)\DB\YB(D\DB\YB)^{2} \nonumber  \\
&&+2ZDXDZ\DB\YB D\DB\YB\DB\ZB-Z(2ZDX+YDY)\DB\YB D\DB\YB D\DB\ZB \nonumber \\
&&+ZDYDZ\DB\YB\DB\ZB D\DB\ZB-Z^{2}DY\DB\YB(D\DB\ZB)^{2}
-Z(XY-Z)(D\DB\YB)^{3} \nonumber \\
&&-Z(YDY-XDZ-ZDX)(D\DB\YB)^{2}\DB\ZB-Z(Y^{2}+XZ)(D\DB\YB)^{2}D\DB\ZB
\nonumber \\
&&-Z(2YDZ+ZDY)D\DB\YB\DB\ZB D\DB\ZB-2YZ^{2}D\DB\YB(D\DB\ZB)^{2}
\nonumber  \\
&&-2Z^{2}DZ\DB\ZB(D\DB\ZB)^{2}-Z^{3}(D\DB\ZB)^{3}.
\ENN
\subsection{Poisson brackets between  $W$ currents}
In this subsection we show that the $W$ supercurrents constructed above
indeed close among themselves under the Poisson bracket defined by
\EE
\left\{\Phi^{i}(Z_{1}),\DB{\bar \Phi}^{\jb}(Z_{2})\right\}
=\delta^{ij}\frac{4\sqrt{-1}}{4\pi}\frac{\th_{12}}{Z_{12}} \label{Poisson}.
\EN
%Regard the right hand side of (\ref{Poisson}) as  a kind of
%Sato's hyperfunction.
We checked that the energy-momentum tensor generates  a center-less
$N=2$ Virasoro algebra under the Poisson bracket.Thus  the structure of
the super $W$ algebra is much  simpler than that of the super Toda theory.
To compute the Poisson brackets for $W$ currents one only needs the bracket
for  spin 1 currents $J(i)$ s by virtue of  the decomposition (\ref{split2}).
It can be shown that (\ref{Poisson}) is equivalent to
\EE
\left\{a(i)(Z_{1}),\DB{\bar b}(\jb)(Z_{2})\right\}
=\delta^{ij}\frac{4\sqrt{-1}}{4\pi}\frac{\th_{12}}{Z_{12}} \label{Poisson2}.
\EN
Then this  induces the following commutation relation
\EEN
&&-\frac{4\pi}{8} \frac{1}{2(k+n+1)} \left\{J(i)(Z_1),J(j)(Z_2)\right\}
\nonumber   \\
&=&\delta^{ij}\left[ \frac{\th_{12}\thb_{12}}{(Z_{12})^{2}}
+\frac{\sqrt{-1}}{4} \frac{\th_{12}}{Z_{12}} D
+\frac{\sqrt{-1}}{4} \frac{\thb_{12}}{Z_{12}} \DB
+\frac{\th_{12}\thb_{12}}{Z_{12}} \pa \right] J(i)(Z_{2}).
\ENN
Notice that the level $k$ of the model appears only as a normalization
constant of the Poisson bracket.\\
$\underline {CP_2 \ {\rm model}} $ \\
In this case we must show that $\big(W^{(1)},W^{(2)}\big)$ form
a closed algebra under the Poisson bracket.
The results are as follows.
\EEN
&&-\frac{4\pi}{8}\frac{1}{2(k+3)}\left\{W^{(1)}(Z_1),W^{(2)}(Z_2)\right\}
\nonumber  \\
&=&\left[2\frac{\th_{12}\thb_{12}}{(Z_{12})^{2}}
+\frac{\sqrt{-1}}{4}\frac{\th_{12}}{Z_{12}}D
+\frac{\sqrt{-1}}{4}\frac{\thb_{12}}{Z_{12}}\DB
+\frac{\th_{12}\thb_{12}}{Z_{12}}\pa \right]W^{(2)}(Z_2)
\ENN
\EEN
&&-\frac{4\pi}{8}\frac{1}{2(k+3)}\left\{W^{(2)}(Z_1),W^{(2)}(Z_2)\right\}=
+\SATB W^{(1)}W^{(2)}(Z_2)  \nonumber  \\
&&+\SATC \left[W^{(1)}DW^{(2)}-W^{(2)}DW^{(1)}\right](Z_2)
+\SATD \left[W^{(1)}\DB W^{(2)}-W^{(2)}\DB W^{(1)}\right](Z_2)\nonumber \\
&&+\frac{\sqrt{-1}}{4} \SATA\left[\DB W^{(2)}DW^{(1)}-\DB W^{(1)}DW^{(2)}
-(D\DB +\DB D)W^{(2)}\cdot W^{(1)}\right](Z_{2})
\ENN
$\underline {CP_3 \ {\rm model}}$\\
In this case we see that $\big(W^{(1)},W^{(2)},W^{(3)}\big)$  close among
themselves  under the Poisson bracket,
\EEN
&&-\frac{4\pi}{8}\frac{1}{2(k+4)}\left\{W^{(1)}(Z_1),W^{(2)}(Z_2)\right\}
\nonumber  \\
&=&\left[2\frac{\th_{12}\thb_{12}}{(Z_{12})^{2}}
+\frac{\sqrt{-1}}{4}\frac{\th_{12}}{Z_{12}}D
+\frac{\sqrt{-1}}{4}\frac{\thb_{12}}{Z_{12}}\DB
+\frac{\th_{12}\thb_{12}}{Z_{12}}\pa \right]W^{(2)}(Z_2)
\ENN
\EEN
&&-\frac{4\pi}{8}\frac{1}{2(k+4)}\left\{W^{(1)}(Z_1),W^{(3)}(Z_2)\right\}
\nonumber  \\
&=&\left[3\frac{\th_{12}\thb_{12}}{(Z_{12})^{2}}
+\frac{\sqrt{-1}}{4}\frac{\th_{12}}{Z_{12}}D
+\frac{\sqrt{-1}}{4}\frac{\thb_{12}}{Z_{12}}\DB
+\frac{\th_{12}\thb_{12}}{Z_{12}}\pa \right]W^{(3)}(Z_2)
\ENN
\EEN
&&-\frac{4\pi}{8}\frac{1}{2(k+4)}\left\{W^{(2)}(Z_1),W^{(2)}(Z_2)\right\}=
\SATB \left[W^{(1)}W^{(2)}+3W^{(3)}\right](Z_2) \nonumber \\
&+&\SATC \left[W^{(1)}DW^{(2)}-W^{(2)}DW^{(1)}+DW^{(3)}\right](Z_2)
 \nonumber \\
&+&\SATD \left[W^{(1)}\DB W^{(2)}-W^{(2)}\DB W^{(1)}+3\DB W^{(3)}
\right](Z_2)  \\
&+&\frac{\sqrt{-1}}{4} \SATA\left[\DB W^{(2)}DW^{(1)}-\DB W^{(1)}DW^{(2)}
 -(D\DB +\DB D)W^{(2)}W^{(1)}-2(D\DB+\DB D)W^{(3)}\right](Z_2) \nonumber
\ENN
\EEN
&&-\frac{4\pi}{8}\frac{1}{2(k+4)}\left\{W^{(2)}(Z_1),W^{(3)}(Z_2)\right\}=
\SATB 2W^{(1)}W^{(3)}(Z_2) \nonumber \\
&+&\SATC \left[W^{(1)}DW^{(3)}-W^{(3)}DW^{(1)}\right](Z_2)
+\SATD\left[W^{(1)}\DB W^{(3)}-W^{(3)}\DB W^{(1)}\right](Z_2) \nonumber \\
&&+\frac{\sqrt{-1}}{4}\SATA \left[DW^{(3)}\DB W^{(1)}-\DB W^{(3)}DW^{(1)}
-(D\DB +\DB D)(W^{(1)}W^{(3)})\right](Z_2)
\ENN
\EEN
&&-\frac{4\pi}{8}\frac{1}{2(k+4)}\left\{W^{(3)}(Z_1),W^{(3)}(Z_3)\right\}=
\SATB W^{(2)}W^{(3)}(Z_2) \nonumber \\
&+&\SATC \left[W^{(2)}DW^{(3)}-W^{(3)}DW^{(2)}\right](Z_2)
+\SATD\left[W^{(2)}\DB W^{(3)}-W^{(3)}\DB W^{(2)}\right](Z_2) \nonumber \\
&&+\frac{\sqrt{-1}}{4}\SATA \left[DW^{(3)}\DB W^{(2)}-\DB W^{(3)}DW^{(2)}
-(D\DB +\DB D)W^{(3)}\cdot W^{(2)}\right](Z_2).
\ENN
%
%
%
%n%
%
%
%
%n
%
We can easily evaluate  the Poisson brackets for the other
$CP_n$ models by the same method.\\
To conclude this section we make two remarks.
Firstly these algebras depend on the level $k$ only through
 the normalization of the brackets.
This is in marked contrast with the super Toda case where the central extenison
term appears even at the classical level.
Hence it is much more important to know the quantum regularization of these
$W$ currents in the Landau-Ginzburg model and compute
the operator product expansions.
Secondly it should be noted that while we  used the superfields
$(a(1),a(2),\ldots ,a(n))$ to express the equation of motion
and Poisson bracket, we cannot use them in quantum theory because of
the nontrivial Jacobian comes out \cite{Ge2} \cite{CeVa}
\EE
{\rm det}\frac{\pa \Phi^{i}}{\pa a(j)}=\prod_{ l_1< l_2}(a(l_1)-a(l_2)).
\EN
%
%
%
%n%nn
%
%
%
%
\section{Discussion}
In this article we investigated the classical aspect of $N=2$ super $W$
symmetry in Landau-Ginzburg models after clarifying the meaning of the
half-twist operation.
 In the half-twisted Landau-Ginzburg model,
$N=2$ super $W$ currents are realized as BRS cohomology classes
in the field space. Remarkable property which we have observed is
that these $W$ currents generate a ring isomorphic to the chiral ring
so long as  the classical product is concerned.
In general normal ordered product in CFT is known to   be neither
commutative nor associative.
Thus it is interestingwould to know the quantum regulalizations to
$N=2$ super $W$ currents discussed here and look into
their normal ordered product structure.
For example in the case of $N=2$ minimal model of level $k$,
$J^{k+1}$ is BRS exact by the classical equation of motion (\ref{null}).
When quantum corrections are made this may repersent the
existence of the singular vector in the vacuum Verma module.
\footnote{We thank  Y. Yamada for pointing out this. Also see \cite{FeMaVaWa}.}
%It would be interesting if one can predict the existence of singular vectors
% in other multi-variable Landau-Ginzburg models in this way.
Our method would be used for analyse the $W$ symmetry of other level 1 coset
 (for example $SO(2n)/U(n)$) models.
Finally the $CP_{n}$ coset models have quantum level description in terms of
the Landau-Ginzburg models \cite{LeVaWa} \cite{Ge2}
and the classical level description by super Toda theories  \cite{NoMo}
\cite{KoMoNo} \cite{EHo}. In these models it is important
to elucidate the Landau-Ginzburg $/$ Toda correspondence \cite{LiMa}. \\
\vspace{0.2cm}
\noindent{\em acknowledgement.}
The author  wishes to  thank  S.-K. Yang  for discussion and
reading the manuscript. The author also wishes to thank K. Ito
for many things. \par

\appendix
\section{Appendix}
\subsection{Notation of $N=2$ superfomalism}

First we collect  some formulas of $N=2$ superformalism used in this paper.
Our convention is that of Wess-Bagger \cite{WeBa} reduced to 2 dimensions
$(x^{0},x^{1},x^{2},x^{3})\rightarrow (x^{0},x^{3})$ \cite{W1}.
Supercovariant differentials are given by
\EE
D_{\pm}= \frac{\partial}{\partial\theta^{\pm}}-2\sqrt{-1}{\bar \theta^{\pm}}
\partial_{\pm},\ \
{\bar D_{\pm}}=-\frac{\partial}{\partial{\bar \theta^{\pm}}}
+2\sqrt{-1}\theta^{\pm}\partial_{\pm}
\EN
and supercharges are
\EE
Q_{\pm}=\frac{\pa}{\pa \th^{\pm}}+2\sqrt{-1}{\bar \th}^{\pm}\pa_{\pm},
\ \ {\bar Q}_{\pm}=-\frac{\pa}{\pa{\bar \th}^{\pm}}
-2\sqrt{-1}\th^{\pm}\pa_{\pm}
\EN.
Right-moving (+)and left-moving($-$) coordinates are defined as
\EE
{\bar \theta^{\pm}}=(\theta^{\pm})^{*},x^{\pm}=x^{0}\pm x^{3}
\EN
Supertransformations for general superfield $A$ are defined by
\EE
\delta_{\xi}A=(\xi^{+}Q_{+}+\xi^{-}Q_{-}-{\bar \xi}^{+}{\bar Q}_{+}
-{\bar \xi}^{-}Q_{-})A.
\EN
Chiral superfield $\Phi $ is defined as
\EE
{\bar D}_{\pm}\Phi=0
\EN
and is expanded as follows,
\EE
%\Phi =\Phi(y^{-},y^{+},\th^{-},\th^{+})
\Phi    =\phi(y^{-},y^{+})+\sqrt{2}\th^{-}\psi_{-}(y^{-},y^{+})
    +\sqrt{2}\th^{+}\psi_{+}(y^{-},y^{+})+2\th^{+}\th^{-}F(y^{-},y^{+})
\EN
\EE
y^{\pm}=x^{\pm}+2\sqrt{-1}{\bar \th}^{\pm}\th^{\pm}
\EN
We also define anti chiral superfield $ {\bar \Phi}$
\EE
D_{+}{\bar \Phi}=0, D_{-}{\bar \Phi}=0
\EN
\EE
{\bar \Phi}={\bar \phi}({\bar y}^{-},{\bar y}^{+})
-\sqrt{2}{\bar \th}^{-}{\bar \psi}_{-}({\bar y}^{-},{\bar y}^{+})
-\sqrt{2}{\bar \th}^{+}{\bar \psi}_{+}({\bar y}^{-},{\bar y}^{+})
+2{\bar \th}^{-}{\bar \th}^{+}{\bar F}({\bar y}^{-},{\bar y}^{+}).
\EN
\EE
{\bar y}^{\pm}=x^{\pm}+2\sqrt{-1}\th^{\pm}{\bar \th}^{\pm}
\EN
We finally give the super transformations of component fields using the
equation of motion of $F$ field.
\EEN
\delta \phi^{i}&=&\sqrt{2}(\xi^{-}\psi_{-}^{i}+\xi^{+}\psi_{+}^{i}),
\delta{\bar \phi}^{\bar i}=-\sqrt{2}({\bar \xi}^{-}{\bar \psi}_{-}^{\bar i}
+{\bar \xi}^{+}{\bar \psi}_{+}^{\bar i}) \\
\delta\psi_{-}^{i}&=&\sqrt{2}(-\frac{1}{4}\xi^{+}\pa_{\bar i}{\bar W}
-2\sqrt{-1}{\bar \xi}^{-}\pa_{-}\phi^{i}),
\delta{\bar \psi}_{-}^{\bar i}=\sqrt{2}(-\frac{1}{4} {\bar \xi}^{+}\pa_{i}W
+2\sqrt{-1}\xi^{-}\pa_{-}{\bar \phi}^{\bar i}) \\
\delta\psi_{+}^{i}&=&\sqrt{2}(\frac{1}{4}\xi^{-}\pa_{\bar i}{\bar W }
-2\sqrt{-1}{\bar \xi}^{+}\pa_{+}\phi^{i}),
\delta{\bar \psi}^{\bar i}=\sqrt{2}(\frac{1}{4} {\bar \xi}^{-}\pa_{i}W
+2\sqrt{-1}\xi^{+}{\bar \phi}^{\bar i})
\ENN
\subsection{Operator product expansion formulas}
We give operator product expansion formulas
in Landau-Ginzburg model which may be useful in  discussing the free field
realization of $N=2$ algebra in terms of the Landau-Ginzburg superfilds.
The left moving super coordinates are
\EEN
\th &=& \th^{-}, \ \  \thb=\thb^{-} ,\ \ z_{12}=x^{-}_{1}-x^{-}_{2}\nonumber \\
Z_{12} &=& z_{12}+2\sqrt{-1}(\th_{1}\thb_{2}-\th_{2}\thb_{1}),
 \ \ \ \th_{12}=\th_{1}-\th_{2} ,\thb_{12}=\thb_{1}-\thb_{2} \nonumber    \\
\Phi(Z_{1}){\bar \Phi}(Z_{2}) &=& \frac{1}{4\pi}
\log(Z_{12}-2\sqrt{-1} \th_{12}\thb_{12})  \nonumber \\
{\bar \Phi}(Z_{1})\Phi(Z_{2}) &=& \frac{1}{4\pi}
\log(Z_{12}+2\sqrt{-1} \th_{12}\thb_{12})  \nonumber \\
\Phi(Z_{1})\DB{\bar \Phi}(Z_{2}) &=& \DB{\bar \Phi}(Z_{1})\Phi(Z_{2})
=\frac{4\sqrt{-1}}{4\pi}\frac{\th_{12}}{Z_{12}}\nonumber \\
 D\Phi(Z_{1}){\bar \Phi}(Z_{2}) &=& {\bar \Phi}(Z_{1})D\Phi(Z_{2})
=-\frac{4\sqrt{-1}}{4\pi}\frac{\thb_{12}}{Z_{12}}   \nonumber \\
D\Phi(Z_{1})\DB{\bar \Phi}(Z_{2})&=&\frac{4\sqrt{-1}}{4\pi}
\frac{1}{Z_{12}+2\sqrt{-1}\th_{12}\thb_{12}} \nonumber \\
\DB{\bar \Phi}(Z_{1})D\Phi(Z_{2})&=&\frac{4\sqrt{-1}}{4\pi}
\frac{1}{Z_{12}-2\sqrt{-1}\th_{12}\thb_{12}}
\ENN
The Taylor expansion for $N=2$ superfield is as follows,
\EE
A(z_{1},\th_{1},\thb_{1})=
\sum \frac{1}{k!}(Z_{12})^{k}\left [ 1+\th_{12}D_{2}-\thb_{12}\DB_{2}
+\frac{1}{2}\th_{12}\thb_{12}(D_{2}\DB_{2}-\DB_{2}D_{2}) \right ]\pa^{k}
A(z_{2},\th_{2},\thb_{2}).
\EN
The energy-momentum tensor is given as
\EE
J = \frac{1}{2}(1-\alpha_{i})D\Phi^{i}\DB{\bar \Phi}^{\ib}
-\frac{\alpha_{i}}{2}\Phi^{i}D\DB{\bar \Phi}^{\ib}
 =  \frac{1}{2}D\Phi^{i}\DB{\bar \Phi}^{\ib}
-\frac{\alpha_{i}}{2}D(\Phi^{i}\DB{\bar \Phi}^{\ib})
\EN
The normalized energy-momentum tensor is
 ${\displaystyle T= -\frac{4\pi}{8}J}$, and if A is a primary field with
$(L_{0},J_{0})=(h,q)$, then
\EEN
T(Z_{1})A(Z_{2}) &=& \left[\frac{\sqrt{-1}}{4}q\frac{1}{Z_{12}}
+h\frac{\th_{12}\thb_{12}}{(Z_{12})^{2}}
+\frac{\sqrt{-1}}{4}\frac{\th_{12}}{Z_{12}}D
+\frac{\sqrt{-1}}{4}\frac{\thb_{12}}{Z_{12}}\DB
+\frac{\th_{12}\thb_{12}}{Z_{12}}\pa \right]A(Z_{2})\nonumber \\
T(Z_{1})T(Z_{2}) &=& -\frac{1}{16} {\hat c}\frac{1}{(Z_{12})^{2}} \nonumber \\
&+& \left[\frac{\th_{12}\thb_{12}}{(Z_{12})^{2}}
+\frac{\sqrt{-1}}{4}\frac{\th_{12}}{Z_{12}}D
+\frac{\sqrt{-1}}{4}\frac{\thb_{12}}{Z_{12}}\DB
+\frac{\th_{12}\thb_{12}}{Z_{12}}\pa \right]T(Z_{2}) \nonumber
\ENN
In particular for the fundamental chiral fields,
\EEN
T(Z_{1})\Phi^{i}(Z_{2}) &=&  \left[\frac{\sqrt{-1}}{4}
\alpha_{i}\frac{1}{Z_{12}}
+\frac{\alpha_{i}}{2}\frac{\th_{12}\thb_{12}}{(Z_{12})^{2}}
+\frac{\sqrt{-1}}{4}\frac{\th_{12}}{Z_{12}}D
+\frac{\th_{12}\thb_{12}}{Z_{12}}\pa \right]\Phi^{i}(Z_{2}) \nonumber \\
T(Z_{1})\DB{\bar \Phi}^{\ib}(Z_{2})
 &=& \left[\frac{\sqrt{-1}}{4}(1-\alpha_{i})\frac{1}{Z_{12}}
+\frac{1}{2}(1-\alpha_{i})\frac{\th_{12}\thb_{12}}{(Z_{12})^{2}}
+\frac{\sqrt{-1}}{4}\frac{\th_{12}}{Z_{12}}D
+\frac{\th_{12}\thb_{12}}{Z_{12}}\pa \right]\DB{\bar \Phi}^{\ib}(Z_{2})
\nonumber
\ENN
\section{Ring structure of $W$ currents}
Here we give the complete expression of (\ref{n2}).
\EEN
&&(10J)(2W^{(2)}) = \DPB \bigl[-24XDX\DB\XB\DB\YB(D\DB\YB)^{2}
+6X^{2}D\DB\XB\DB\YB(D\DB\YB)^{2} \nonumber  \\
&& 4XY\DB\YB(D\DB\YB)^{3}+(D\DB\XB)^{3}\DB\YB
 +\DB\XB(D\DB\XB)^{2}D\DB\YB \bigl]
\ENN
\EEN
&&(10J)^{4}+(2W^{(2)})^{2} = \DPB \bigl[
96X^{3}DX\DB\XB\DB\YB(D\DB\YB)^{3}
-24X^{4}D\DB\XB\DB\YB(D\DB\YB)^{3}  \nonumber     \\
&&-12X^{3}Y\DB\YB(D\DB\YB)^{4}
-96XDX\DB\XB(D\DB\XB)^{2}\DB\YB D\DB\YB \nonumber   \\
&&+12X^{2}(D\DB\XB)^{3}\DB\YB D\DB\YB
+8X^{2}(D\DB\XB)^{2}(D\DB\YB)^{2}  \nonumber     \\
&&+192XDY\DB\XB D\DB\XB\DB\YB(D\DB\YB)^{2}
+24XY(D\DB\XB)^{2}\DB\YB(D\DB\YB)^{2} \nonumber    \\
&&+64XY\DB\XB D\DB\XB(D\DB\YB)^{3}
+192YDY\DB\XB\DB\YB(D\DB\YB)^{3} \nonumber   \\
&&+32Y^{2}\DB\XB(D\DB\YB)^{4}
 +2Y^{2}\DB\XB(D\DB\YB)^{4} \bigl]
\ENN
The right hand side of (\ref{n3}) is given by
\EEN
(14J)(3W^{(3)})&=&\DPB\bigl[
-9X^{8}\DB\YB(D\DB\YB)^{4}+144X^{3}DX\DB\XB D\DB\XB\DB\YB(D\DB\YB)^{2}
\nonumber  \\
&&-6X^{4}(D\DB\XB)^{2}\DB\YB(D\DB\YB)^{2}
-18X^{4}\DB\XB D\DB\XB(D\DB\YB)^{3}  \nonumber  \\
&&+48X^{3}DY\DB\XB\DB\YB(D\DB\YB)^{3}-24X^{3}YD\DB\XB\DB\YB(D\DB\YB)^{3}
\nonumber  \\
&&-(D\DB\XB)^{4}\DB\YB-6\DB\XB(D\DB\XB)^{3}D\DB\YB \bigl]
\ENN
\EEN
&&(14J)^{6}+(3W^{(3)})^{2} = \DPB \bigl[
-8640X^{11}DX\DB\XB\DB\YB(D\DB\YB)^{5}+1512X^{12}\DB\XB(D\DB\YB)^{6}
\nonumber \\
&&+X^{12}\DB\XB(D\DB\YB)^{6}
-648X^{7}DX\DB\XB(D\DB\XB)^{2}\DB\YB(D\DB\YB)^{3} \nonumber \\
&&+68X^{8}(D\DB\XB)^{3}\DB\YB(D\DB\YB)^{3}
+36X^{8}\DB\XB(D\DB\XB)^{2}(D\DB\YB)^{4} \nonumber  \\
&&-14832X^{7}DY\DB\XB D\DB\XB\DB\YB(D\DB\YB)^{4}
-1152X^{7}Y(D\DB\XB)^{2}\DB\YB(D\DB\YB)^{4} \nonumber  \\
&&-2916X^{7}Y\DB\XB D\DB\XB(D\DB\YB)^{5}
-11664X^{6}YDY\DB\XB\DB\YB(D\DB\YB)^{5} \nonumber   \\
&&-1458X^{6}Y^{2}\DB\XB(D\DB\YB)^{6}
-576X^{3}DX\DB\XB(D\DB\XB)^{4}\DB\YB D\DB\YB \nonumber  \\
&&+24X^{4}(D\DB\XB)^{5}\DB\YB D\DB\YB
+18X^{4}\DB\XB(D\DB\XB)^{4}(D\DB\YB)^{2} \nonumber  \\
&& -3288X^{2}YDX\DB\XB(D\DB\XB)^{3}\DB\YB(D\DB\YB)^{2}
-120X^{3}DY\DB\XB(D\DB\XB)^{3}\DB\YB(D\DB\YB)^{2}\nonumber  \\
&&+144X^{3}Y(D\DB\XB)^{4}\DB\YB(D\DB\YB)^{2}
-16X^{3}Y\DB\XB(D\DB\XB)^{3}(D\DB\YB)^{3}\nonumber  \\
&&+9672XY^{2}DX\DB\XB(D\DB\XB)^{2}\DB\YB(D\DB\YB)^{3}
+548X^{2}Y^{2}(D\DB\XB)^{3}\DB\YB(D\DB\YB)^{3} \nonumber  \\
&&+2430X^{2}Y^{2}\DB\XB(D\DB\XB)^{2}(D\DB\YB)^{4}
+19440XY^{2}DY\DB\XB D\DB\XB\DB\YB(D\DB\YB)^{4}\nonumber  \\
&&+2916XY^{3}XY^{3}\DB\XB D\DB\XB(D\DB\YB)^{5}
+11664Y^{3}DY\DB\XB\DB\YB(D\DB\YB)^{5} \nonumber  \\
&&+1458Y^{4}\DB\XB(D\DB\YB)^{6}
+2\DB\XB(D\DB\XB)^{6} \bigl].
\ENN
\section{The computation of $M^{(k)}$}
We first note that the left hand side of (\ref{anti}) can be expanded
as
\EEN
&&\DB{\bar b}(1)(D\DB{\bar b}(1))^{k+2}
-\DB{\bar b}(2)(D\DB{\bar b}(2))^{k+2} \nonumber \\
&&=(a(1)-a(2))N^{(k)}+D(a(1)-a(2))S^{(k)},
\ENN
where $N^{(k)},S^{(k)}$ are elements of ${\bf C}[X,Y,\DB \XB,\DB \YB;D]$.
In the same way we have
\EEN
-2(k+3)(J(1)-J(2))
&=&(a(1)-a(2))\left[D\DB \XB+\frac{1}{2}(k+3)DX\DB \YB\right] \nonumber \\
&+&D(a(1)-a(2))\left[-\frac{1}{2}(k+3)X\DB \YB -(k+2)\DB \XB \right].
\ENN
$S^{(k)}$ can be directly computed as
\EEN
S^{(k)} &=& (k+2)\DB \XB \DB \YB \nonumber \\
&\times & \sum_{l=0}^{k+2}\frac{1}{l+1}{}_{k+2}C_{l}
(a(1)^{l}+a(1)^{l-1}a(2)+\cdots +a(2)^{l})(D\DB \XB)^{k+2-l}(D\DB \YB)^{l}.
\ENN
Finally we can solve $M^{(k)}$ by
\EE
S^{(k)}=\left[\frac{1}{2}(k+3)X\DB \YB+(k+2)\DB \XB \right]M^{(k)}.
\EN

\end{document}